\def\spose#1{\hbox to 0pt{#1\hss}}
\def\lta{\mathrel{\spose{\lower 3pt\hbox{$\mathchar"218$}}
     \raise 2.0pt\hbox{$\mathchar"13C$}}}
\def\gta{\mathrel{\spose{\lower 3pt\hbox{$\mathchar"218$}}
     \raise 2.0pt\hbox{$\mathchar"13E$}}}

\documentstyle[12pt]{article}
\def\reference{\parskip 0pt\par\noindent\hangindent 0.5 truecm}

\begin{document}

\title{The Cloudy Universe}

\author{Mark Walker \& Mark Wardle
} 

\date{}
\maketitle

{\center
Special Research Centre for Theoretical Astrophysics,
School of Physics,\\University of Sydney, NSW 2006\\
M.Walker/M.Wardle@physics.usyd.edu.au\\[3mm]
}

\begin{abstract}

Modelling of Extreme Scattering Events suggests that the Galaxy's
dark matter is an undetected population of cold, AU-sized,
planetary-mass gas clouds. None of the direct observational
constraints on this picture -- thermal/non-thermal emission,
extinction and lensing -- are problematic.
The theoretical situation is less comfortable, but still
satisfactory. Galactic clouds can survive in their current
condition for billions of years, but we do not have a
firm description for either their origin or their evolution
to the present epoch. We hypothesise that the proto-clouds
formed during the quark-hadron phase transition, thereby
introducing the inhomogeneity necessary for compatibility
with light element nucleosynthesis in a purely baryonic
universe. We outline the prospects for directly detecting
the inferred cloud population. The most promising signatures
are cosmic-ray-induced H$\alpha$  emission from clouds in the
solar neighbourhood, optical flashes arising from cloud-cloud
collisions, ultraviolet extinction, and three varieties of
lensing phenomena.

\end{abstract}

{\bf Keywords: dark matter --- galaxies: halos --- ISM: clouds}

\bigskip

\section{Introduction}

This paper traces some lines of thought concerning the nature of dark 
matter which the authors have been pursuing for the last year or so.  
For the most part its content follows the talk given by one of us 
(Mark I) to the July 1998 Meeting of the Astronomical Society of 
Australia in Adelaide, with some extra material reflecting subsequent 
developments.  

We begin with a brief overview of current activity directed towards 
identifying dark matter (\S 2), then focus on a model in 
which the Galactic dark matter is composed of planetary-mass gas 
clouds --- a picture that can be drawn directly from the radio-wave 
lensing of quasars reported by Fiedler et al (1987, 1994).  A lensing 
event (Extreme Scattering Event, or ESE) occurs when a cloud in the 
Galactic halo crosses the line of sight: radio waves are refracted by 
an ionised wind which is evaporated from the neutral cloud by UV
radiation (\S3). The observed event rate implies that the total mass
in clouds is comparable to that of the Galactic dark matter.

This inferred population of $\sim 10^{15}$ cold clouds, each of mass
$\sim 10^{-4}\;{\rm M_\odot}$ and size $\sim$ AU, suprisingly, does not
violate  existing observational constraints (\S 4).  The theoretical issues 
raised by this scenario, discussed in \S 5, are quite basic: how did 
the clouds form, and how have they survived?  In \S6 we consider other 
techniques for detecting the cloud population.  The most 
promising signatures are optical flashes from cloud collisions, 
H$\alpha$ emission from nearby clouds, and various lensing phenomena.  
Finally, the outlook for the model is discussed in \S 7.

\section{Identification of dark matter}

Zwicky (1933) was the first to infer the existence of large quantities
of unseen mass, in studies of the dynamics of clusters of galaxies,
but it was many years before the ``missing mass'', or dark matter,
problem was widely recognised. Van~den~Bergh (1999) reviews the early
history of this subject, while Ashman (1992) gives a thorough account
of the problem in the context of galaxies. It is, in fact, HI galaxy
rotation curves which provide the best evidence we have for the existence
of dark matter; here the dynamical interpretation, and its implication,
is entirely unambiguous. Equally interesting, though less clear-cut, are
the indications that the Universe as a whole is composed principally of
dark matter --- this evidence is discussed extensively by Peebles (1993).

Because stars usually contribute the bulk of the visible mass, at 
least in galaxies, much attention has been given to the possibility 
that dark matter is composed of stars which are of low luminosity.  
This category includes: low-mass main-sequence stars; brown dwarfs; 
old white dwarfs; neutron stars; and black holes --- a diverse group 
which has required a variety of techniques to constrain their total 
mass contribution.  It is beyond the scope of this paper to summarise 
these efforts; good reviews are given by Trimble (1987), Ashman (1992) 
and Carr (1994).  However, a unifying feature of this group is that 
they are all rather dense objects and so constitute strong 
gravitational lenses even when they are in the halo of our Galaxy.  As 
pointed out by Paczy\'nski (1986), this enables indirect searches for 
such objects via photometric monitoring of millions of Magellanic 
cloud stars, and such gravitational lensing events have now clearly 
been detected (Alcock et al 1997).  It is not currently known whether 
these lensing events are caused by dark matter, or by known stellar 
populations associated with the Galaxy and the Magellanic Clouds (Sahu 
1994).

If it turns out that stars don't fit the bill, what might the dark
matter be composed of? The lack of success of very deep searches
for previously uncatalogued material, conducted with the most modern
instrumentation, across the whole range of the observable spectrum,
promoted the suspicion that dark matter has {\it no\/} electromagnetic
interaction. In this case one imagines the dark matter to be composed
of massive neutrinos, for example, or perhaps an elementary particle
which has not yet been detected? This idea was strengthened by detailed
computations of the abundances of light elements which
result from primordial nucleosynthesis in a hot Big Bang cosmology.
These calculations, when compared to the observed abundances,
suggest that only a small fraction of the closure density is in
the form of baryonic (ordinary) material --- see the review by
Schramm \& Turner (1998). This picture of weakly interacting dark
matter readily lends itself to numerical simulations of the growth
of structure (galaxies, clusters, superclusters), from an initially
near-homogeneous universe. These simulations demonstrate that
dynamically `cold' dark matter can reproduce
the observed structural characteristics of the Universe, at least
approximately, while simultaneously remaining consistent with
measurements of the Cosmic Microwave Background (CMB) anisotropies.
`Hot' dark matter gives a much poorer representation of the
observed structure (Davis et al 1985), suggesting that the dark
matter is not composed of neutrinos, but is a type of particle
which has yet to be detected. These results have given impetus to
a variety of experiments designed to detect individual dark matter
particles, or their decay products --- see the many contributions
to Spooner (1997).

Contrary to these general trends in the field, Pfenniger, Combes
\& Martinet (1994: PCM94 hereafter) advanced the view that the
dark matter associated with galaxies might be composed of cold
gas clouds.  PCM94 motivated this suggestion with arguments
related to galaxy dynamics and evolution, emphasising the
astrophysical appeal of dark matter in this form, and noting
that if the temperature of the cold gas were close to that of
the CMB then the clouds would be very difficult to detect (see also
Combes \& Pfenniger 1997). PCM94 additionally proposed that the
cold gas should have a fractal structure -- an idea which was
developed in Pfenniger \& Combes (1994) -- and that it
should be distributed in a thin disk. However, neither of these
are essential features for a dark matter model based on cold gas,
and one can equally well imagine a spherical halo of individual or
clustered clouds (de~Paolis et al 1995; Gerhard \& Silk 1996).
These quasi-spherical distributions of dark matter are more palatable
to most dynamicists than highly flattened distributions, but even so
there remain plenty of contentious issues relating to the physics of
the putative cold gas clouds; these are addressed in \S\S4,5; first
we turn to the main observational evidence for their existence.

\section{Extreme Scattering Events}
``Extreme Scattering Events'' (ESEs) were discovered more than a
decade ago (Fiedler et al 1987), during a multi-year program to
monitor the positions and radio fluxes of a large number of compact
quasars. The events themselves amount to large, frequency-dependent
variations in the received radio signal, lasting for a month or two.
These variations were immediately recognised as a Galactic lensing
phenomenon (Fiedler et al 1987; Romani, Blandford \& Cordes 1987),
rather than being attributable to intrinsic changes in the source.
Each lens is in this case just a localised over-density of ionised
gas, which refracts the radio-waves, and flux variations occur when
a lens crosses the line-of-sight. The lenses are inferred to be a
few AU in radius. In the decade following their discovery there was
little progress in understanding the physical nature of these lenses;
the possibility of generating high electron densities within interstellar
shock waves had been explored (e.g. Clegg, Chernoff \& Cordes 1988),
but without much success in reproducing the observed light-curves. A
particular difficulty with the physics of the early models is that the
pressure of the ionised gas is a thousand times larger than that of
the diffuse interstellar medium, so the latter cannot confine the
former and such a lens ought to explode, lasting only a year or so.
In order to avoid this, while still maintaining a static lens, one
is forced to resort to some contrivance with magnetic fields or
highly elongated lenses.

Motivated by a desire for a simpler explanation, we investigated a 
specific physical picture for the lenses in which the hot gas is not 
static but, rather, forms a continuous outflowing wind (Walker \& 
Wardle 1998a).  In turn this implies a reservoir of neutral material 
at the base of the wind, and to ensure longevity of the lens we 
assumed this (cold) neutral cloud to be in hydrostatic equilibrium, 
with its thermal pressure balanced by self-gravity.  This turns out to 
offer a good model for the ESEs: a cold, neutral gas cloud in the 
Galactic halo is expected to develop a photo-evaporated wind as a 
consequence of the ionising radiation field arising from hot stars in 
the Galactic disk (Dyson 1968; see also Henriksen \& Widrow 1995); the 
intensity of this radiation naturally generates the requisite density 
of ionised gas.  Moreover the computed light curves, both at low and 
high radio frequencies, readily reproduce those of the archetypal ESE 
in the source 0954+658.  Indeed as a model for the ESE phenomenon 
there seem to be no real difficulties with this picture.

However, it follows from the ESE rate (Fiedler et al 1994), in the
context of this physical model, that the neutral gas clouds constitute
a large fraction of the mass of the Galaxy. In other words, since they
are not present in our current inventories of visible matter, they are
a major component of dark matter. This conclusion immediately prompts
concerns: is this model consistent with other data? What about the
constraints from Big Bang nucleosynthesis? How could such clouds
form? How can they survive for so long? The next sections address
these issues.

\section{Compatibility with other data}

In order that the mass in neutral gas clouds not exceed the 
dynamically determined value (i.e that deduced from the rotation 
curve) for the Galaxy, the individual cloud masses must be very small: 
$M<10^{-3}\;{\rm M_\odot}$.  Assuming cloud masses comparable to this 
limit, and radii of a few AU implies that they have internal densities 
$n\sim10^{12}\;{\rm cm^{-3}}$, very dense in comparison with any other 
component of the interstellar medium.  Three-body reactions (e.g.  
${\rm 3\,H\rightarrow H+H_2}$; Palla, Salpeter \& Stahler 1983) 
consequently proceed rapidly and the hydrogen is expected to be in 
molecular form.

Unfortunately there is very little else that one can deduce {\it a 
priori\/} about the putative neutral gas clouds, and it is necessary 
to contemplate a broad range of possibilities.  With this in mind we 
now address the issue of detectability of the clouds themselves: ought 
they to be manifest in other ways?

\subsection{Thermal emission}
The most obvious expectation of the model is that there ought to be thermal
emission from these dense clouds, and the Galactic population as a whole
should therefore introduce an extra background of thermal radiation. The
expected background spectrum is dictated by the temperature, density and
composition of the clouds; we know that the temperature is low, and the density
high, but the cloud chemistry is poorly constrained. We can nevertheless arrive
at some crude estimates as follows.

Let us approximate any line emission as optically thick within the 
Doppler core, of velocity width $c_s$, implying a brightness 
temperature equal to the kinetic temperature ($T$) of the cloud, with 
negligible optical depth (and therefore brightness temperature) 
outside this region.  When viewed as a background -- i.e.  without 
resolving the individual clouds -- the observed brightness temperature 
contribution at line centre is then $\Delta T_B \simeq f\,T 
(c_s/\sigma)$, where $f$ is the fraction of the sky covered by clouds, 
and $\sigma$ is their velocity dispersion.  The quasar monitoring data 
(Fiedler et al 1994) suggest that $f\sim5\times10^{-3}$, while 
$c_s\ll\sigma\sim150\;{\rm km\,s^{-1}}$, and we see that
$\Delta T_B$ is very small.  If the 
instrument we are using cannot resolve the line (i.e.  $2\times$~the 
channel width is greater than $\sigma$), then the recorded brightness 
temperature perturbation is even smaller: $\Delta T_B\sim{\cal 
R}T^{3/2}$~nK, in a single spectral channel, where ${\cal R}$ is the 
spectral resolving power and $T$ is in Kelvin.  Inserting parameters 
appropriate to the COBE FIRAS instrument, which had brightness 
temperature sensitivity of order 0.1~mK, and ${\cal R}\sim10^2$ (Fixsen 
et al 1994) we see that FIRAS would not have detected the line unless 
$T\gta100$~K. The foregoing discussion is straightforwardly extended to a 
small number of spectral lines.  While these estimates are rather 
crude they are not specific to any particular coolant: the key point 
is that the microwave/FIR data give only weak limits on 
possible cloud temperatures if there is negligible continuum opacity.

The limits on broad-band radiation are much more restrictive. COBE
FIRAS data revealed a Galactic component of continuum emission
which could be interpreted in terms of dust with temperatures
in the range 4--7~K, and mean optical depth
$\langle\tau\rangle\sim10^{-4}$ (at $\lambda = 0.33$~mm)
at high latitude (Reach et al 1995; see Lagache et al 1998 for
an alternative interpretation).
In the cold cloud model $\langle\tau\rangle=f\,\tau$,
where $\tau$ is the optical depth of a single cloud. Requiring the
thermal continuum emission from any cloud population to be smaller than
the COBE measurements, we then demand $\tau<2\times10^{-2}$ if $T=7$~K,
say. (Using the smaller value of $f$, i.e. $f_{coll}$, derived in
\S5.4, this limit relaxes to $\tau<0.5$.)
In turn, this limit on the optical depth implies that the clouds
contain essentially no dispersed dust. This is expected if their
composition reflects the standard Big Bang abundances, and likely
even if the clouds do contain metals, because any dust grains will
settle into the centre of the cloud, forming a ``dirty snowball'' there.
In \S5.3 we shall present theoretical arguments that there {\it must\/} in
fact be some continuum opacity in these clouds -- not from dust as
such, but from particles of solid hydrogen. The limit on $\tau$ just
quoted applies equally well to these particles.

\subsection{Non-thermal emission}
The most striking feature of the $\gamma$-ray sky is the
luminous interstellar medium in the Galactic plane. This
emission arises as a consequence of cosmic rays interacting
with the gas, principally by nuclear interactions (i.e.
cosmic-ray protons + target nuclei) leading to pion production,
with subsequent decay ($\pi^0\rightarrow2\gamma$);
relativistic electron bremsstrahlung also contributes (e.g.
Bloemen 1989). If we suppose that the dark matter is simply
cold gas then it too will be luminous in $\gamma$-rays, as
a result of these processes (de Paolis et al 1995).
In principle this provides a strong constraint on the amount of
dark matter in cold gas, but in practice the constraint is quite
weak because the Galactic distribution of cosmic rays is
poorly known; in particular the scale-height of the cosmic ray
disk of the Galaxy is uncertain (Webber, Lee \& Gupta 1992).
In addition the column density of individual dark clouds may be
sufficiently high that they are not entirely transparent to
$\gamma$-rays and cosmic rays. As a result of this freedom in
modelling, the $\gamma$-ray data must be regarded as inconclusive
at present. However, we note that Dixon et al (1998) discovered
an unmodelled Galactic ``halo'' component in the $\gamma$-ray
background, and the simplest explanation for this is that it is
due to unseen (cold, dense) gas.

In addition to cosmic rays interacting within each cloud, we have
already noted (\S3) that UV photons are absorbed near the cloud
surface, and drive a wind therefrom. Radiation from this wind, in
particular the optical/IR line transitions of molecular and atomic
hydrogen, might be detectable. (This is, of course, thermal emission,
but it seems more appropriate to cover it here than in \S4.1 because
the wind is so much hotter than the underlying cloud.) Noting
that the Galactic halo ionising radiation has an intensity of
order $2\times10^6/\pi\;\,{\rm cm^{-2}\,s^{-1}\,sr^{-1}}$
(Dove \& Shull 1994), and that all of the ionising photons
incident upon a cloud will be absorbed, one can estimate
(e.g. Bland-Hawthorn \& Maloney 1997) that each cloud should
have an Emission Measure of ${\rm EM\sim2\;\, pc\,cm^{-6}}$.
Averaging over the population of clouds, which collectively cover
only a small fraction of the sky, then leads to a mean surface
brightness (EM) a factor $f\sim5 \times10^{-3}$ smaller.
This emission therefore contributes a tiny fraction ($\sim$1\%) of
the observed EM at high Galactic latitudes (Reynolds 1992), and the
data do not currently provide strong limits on the proposed cloud
population. This situation could be improved with studies of the
low-level emission-line wings, as the clouds are expected to
have a velocity dispersion ($\sigma\simeq150\;{\rm km\,s^{-1}}$)
which is much greater than the width of the dominant observed
component.

We can also consider the possibility of detecting individual clouds
passing through regions where the radiation field is high, i.e HII
regions, but the prospects seem remote. Here we differ from the
conclusions of Gerhard \& Silk (1996), who considered clouds
having radii some four
orders of magnitude larger (hence solid angles, and ${\rm H\alpha}$
fluxes, eight orders of magnitude larger) than in our model.

\subsection{Extinction events}

If the Galactic dark matter is composed of clouds which cover a fraction
$f\sim5\times10^{-3}$ of the sky then, because the total column
of dark matter is known from dynamics (e.g. Binney \& Tremaine 1987)
to be $\sim100\;{\rm M_\odot}$, one can
immediately infer that the column density of each cloud is
$N\sim10^{24}\;{\rm cm^{-2}}$. If these clouds followed the
same relationship between extinction and column density as the
known molecular gas in the Galaxy, this would correspond to several
hundred magnitudes of visual extinction; in turn this would mean
that 0.5\% of extragalactic stars (e.g. in the LMC) would be
completely extinguished, with these events lasting tens of days.
This phenomenon is readily detectable but has not been reported,
and we are obliged to conclude that the clouds cannot be suffused
with dust (see also \S4.1).

In the UV -- X-ray bands each cloud must be opaque because the
fundamental constituents (H$_2$ and He) themselves provide substantial
opacities in these regions of the spectrum (cf. Combes \& Pfenniger 1997).
The main sources of
opacity are: Rayleigh scattering (near UV); the Lyman and Werner
transitions of H$_2$; photoelectric absorption; and electron
scattering at high energies. Monitoring extragalactic sources in
these wavebands should, therefore, unambiguously reveal the
presence of such a cloud population via the extinction
events they introduce. Insufficient data have been recorded to date
to usefully constrain a cold-cloud population, so this remains an
interesting experiment for the future.

\subsection{Lensing events}

There are several distinct phenomena, relevant to cold clouds,
which fall under the general heading of ``lensing'':
plasma lensing (refraction by ionised gas,
giving rise to ESEs), gas lensing (refraction by neutral gas),
and gravitational lensing (refraction by the gravitational field).
The last of these is widely acknowledged as a tool for investigating
dark matter, and it is appropriate to deal with it first.

\subsubsection{Gravitational lensing}
Searches for gravitational microlensing of stars in the Large Magellanic
Cloud (e.g. Alcock et al 1997) are designed to detect flux increases
of 30\% or more, and consequently these experiments are sensitive
only to strong gravitational lenses, i.e. objects with column density
of $10^4\;{\rm g\,cm^{-2}}$ or greater. This is very much greater
than the column density inferred for the cold clouds ($\sim
10^2\;{\rm g\,cm^{-2}}$; see \S5.4), and so the limits on
low mass MACHOs in the Galactic halo (Alcock et al 1998)
are, by definition, irrelevant, even though they cover the critical
planetary mass range.

Clouds located at greater distances, e.g. at cosmological distances,
may be strong gravitational lenses because the column density
required to make a strong lens is smaller in this case, and we expect
that quasars will be gravitationally microlensed by intervening clouds.
Several authors have argued that quasars are indeed microlensed by
planetary-mass objects, including lensing of quasars by cosmologically
distributed objects (Hawkins 1993, 1996), and microlensing of quasars
which are {\it macro\/}lensed by foreground galaxies (e.g. Schild 1996).
The evidence is equivocal at present, and there is a pressing need to
clarify this situation; clarification could be achieved by monitoring
quasars which are viewed through the halos of low redshift galaxies or
clusters (Walker 1999a; Walker \& Ireland 1995; Tadros, Warren \& Hewett
1998). Notice that if Hawkins
is correct, that all quasars exhibit variability as a consequence of
microlensing by planetary mass objects, then it follows (Press \& Gunn
1973) that a large fraction of the cosmological critical density is in
the form of these lenses. This has implications for cosmic nucleosynthesis
(\S5.1) and, in turn, the origin of the clouds (\S5.2).

\subsubsection{Gas lensing}
The concept of ``gas lensing'' was introduced by Draine (1998), who
pointed out that the refractive index of neutral hydrogen/helium gas
clouds would be interestingly large if they had the mass and radius
proposed by Walker \& Wardle (1998a). More specifically: the refractive
index could be comparable to the angular size of Galactic clouds, implying
that strong focussing/de-focussing might occur. Supposing that no such
effects are manifest in the microlensing monitoring data for the
Magellanic Clouds (cf. Alcock et al 1998), Draine (1998) then used
this concept to restrict the acceptable combinations of cloud masses
and radii, under assumed polytropic equations of state. The principal
difficulty in applying this work is our current lack of information
on the density profile of the individual clouds. In turn this reflects
our primitive understanding of the physics relevant to these clouds
(particularly the heating/cooling balance: \S5.3).

\subsubsection{Plasma lensing}

In \S3 we presented the arguments which lead from the observation of
ESEs to the conclusion that the dark matter is composed of cold clouds.
The ESE phenomenon is, however, not the only piece of evidence which
points to the existence of a large population of dense plasma clouds
--- this population is also revealed by observations of periodic
fringes in the spectra of pulsars (Rickett 1990). These periodicities
were discovered many years before ESEs, and their
interpretation requires similarly dense clouds of ionised gas, a few
AU in radius. In principle these observations of pulsars are much more
informative, in respect of the plasma lenses, than observations of ESEs
in quasars, but this advantage has not yet been exploited. Pulsars
clearly offer an opportunity to greatly advance our understanding of
the ionised gas clouds responsible for ESEs, but this requires
systematic studies of the various multiple imaging phenomena.

\section{Theoretical considerations}

In \S4 we described the immediate observational implications of a
population of cold gas clouds contributing to the dark matter;
it is important to recognise
that none of these considerations excludes the model proposed in
\S3. We now turn to issues which are more theoretical, in the sense
that they relate primarily to the physics of the putative clouds.

\subsection{Cosmology}
Theoretical cosmology is responsible for establishing the idea
that dark matter is principally non-baryonic. This idea rests
on two distinct lines of evidence, relating to primordial
nucleosynthesis and to the formation of structure in the Universe.

When the Universe was only minutes old (redshift $z\sim10^9$),
it passed through a period where nuclear reactions were effective
in building-up the abundances of light elements from the initial
building blocks of protons and neutrons. If the Universe is
assumed to be homogeneous then the abundances of the various
elements can be calculated with some precision, subject only to
the unknown photon/baryon ratio (or equivalently the baryonic
contribution, $\Omega_B$, to the closure parameter $\Omega$).
For a small range of values $0.005\lta\Omega_B\lta0.03$, the
calculated abundances of the light elements are roughly in
accord with the observed abundances, giving confidence in the
model and admitting an estimate of $\Omega_B$ --- see Schramm
\& Turner (1998). When combined with other pieces of evidence
that indicate $0.1\lta\Omega\lta1$ (e.g. Peebles 1993), these
calculations demonstrate that most of the Universe (i.e. the
dark matter) was not in the form of smoothly distributed baryons
at the time of nucleosynthesis. This conclusion is usually
stated more succinctly as a requirement for non-baryonic dark
matter.

The second line of evidence for non-baryonic dark matter relates
to the theory of growth of structure in the Universe. The fact
that the Cosmic Microwave Background (CMB) is smooth to $10^{-5}$~K
on large scales (Smoot et al 1992) tells us that the diffuse baryons
possessed very little large-scale inhomogeneity at the epoch when
electrons recombined with ions to form neutral atoms (at
$z\simeq1500$). But today we see a great deal
of structure: galaxies, clusters and even superclusters of galaxies
appear organised in a network of vast filamentary features.
This proliferation of structure can be explained by the effects of
gravity acting on primordial (adiabatic) density fluctuations
if the dark matter couples to the diffuse baryon-photon fluid
only through gravity. If this condition is not met -- e.g. if the
dark matter is supposed to be in the form of diffuse baryons up
to recombination -- it proves difficult to explain the smoothness
of the CMB, on the one hand, and the highly structured local Universe
on the other (Peebles 1993). A known counterexample
to this statement is the baryon isocurvature model -- see Peebles 1993
-- which invokes isothermal primordial density fluctuations.
Unfortunately there is no fundamental theory for the
origin of isothermal fluctuations, and the results for adiabatic
fluctuations are usually given concomitantly greater weight.

These considerations have led to widespread acceptance of the idea
that the bulk of the dark matter is non-baryonic. However, as
emphasised by the phrasing of the preceding paragraphs, neither
case is watertight. Very possibly the Universe was not homogeneous
at the time of nucleosynthesis -- in particular inhomogeneities
could be introduced during the quark-hadron phase transition
(Applegate \& Hogan 1985) --  and in this case the upper limit
on baryons relaxes to $\Omega_B\lta0.3$ (Kurki-Suonio et al 1990).
If we admit the possibility of isothermal fluctuations as the
origin of present-day large-scale structure there is
then no barrier, from cosmological considerations, to a purely
baryonic universe. A baryonic universe might also be
consistent with adiabatic fluctuations, but in this case we
require the proto-clouds to have sufficiently high density
that they decouple from the CMB well ahead of recombination.

\subsection{Cloud formation}

How might planetary-mass gas clouds form? The answer to this question
depends on quite what mass needs to be explained. For clouds
which lie at the upper end of the planetary range,
a fairly straightforward answer can be given: they could form as the
endpoint of hierarchical fragmentation of larger clouds undergoing
collapse in the early Universe ($z\sim100$). There is a substantial
literature on the topic of hierarchical fragmentation, beginning with
Hoyle's (1953) classic paper, mostly employing spherically symmetric
gas clouds in free-fall, in which ``fragment'' masses are equated with
the Jeans mass for the gas. In most calculations the hierarchy is
assumed to terminate when the cloud becomes optically thick -- at
which point cooling is impeded -- and there are sound reasons why
this usually results in sub-stellar masses for the smallest fragments
(Rees 1976). By the same token, however, the smallest fragments are
limited to masses $M>{\rm a\;few\;\times10^{-3}\;M_\odot}$, so if the
putative cold clouds are smaller than this then hierarchical fragmentation
is not viable as a formation scenario.

There are some hints that the individual clouds may indeed have
very small masses (\S5.4), possibly as small as $10^{-5}\;{\rm M_\odot}$.
Moreover the arguments given in the preceding section suggest that
{\it the dark matter is not formed from baryons which were smoothly
distributed at recombination, or even at nucleosynthesis.\/} The
implication is that the proto-clouds formed in the {\it very\/}
early Universe ($z>10^9$, cf. Hogan 1978, 1993), and have
maintained their identity right up to the present. 
(Two further indications that the clouds predate cosmic
nucleosynthesis are the presence of metals in High Velocity Clouds,
and the hints from quasar variability that the Universe might
contain a critical density of planetary-mass objects --- see
\S5.4 and \S4.4.1, respectively.) We envisage that the proto-clouds
formed during the quark-hadron phase transition
(cf. Applegate \& Hogan 1985), and are thus fossils of that era.
Although this is clearly speculative, it is the most economical
hypothesis available.

\subsection{Cloud stability}

Supposing that we can find a sensible explanation for how the
clouds might have formed, it remains to comprehend the
observed/inferred properties of the clouds at the present
time. A major part of this task is to understand the internal
constitution of the clouds, and in particular their thermal
balance (see also Gerhard \& Silk 1996).

For equilibrium, the power radiated from each cloud must be balanced 
by some heat generated within.  Unlike stars this heat is clearly not 
generated by nuclear fusion; there might plausibly be some exothermic 
chemical reactions occurring (e.g.  ${\rm 
2\,H+\,H_2\rightarrow2\,H_2}$), but the simplest hypothesis for 
Galactic clouds is that heat is deposited by energetic particles.  In 
principle this could involve both photons and cosmic rays (their 
Galactic energy densities are similar); however, given that the clouds 
are transparent to optical photons (\S4.3), but not to cosmic 
rays, it is likely that the cosmic rays dominate.

The local cosmic-ray heating rate for dense interstellar molecular gas 
is $\sim3\times10^{-4}\;{\rm erg\,g^{-1}\,s^{-1}}$ (Cravens \& Dalgarno 
1978), and supposing the cloud temperature to be $T\sim10$~K this 
immediately implies a thermal (Kelvin-Helmholtz) time-scale of order 
$10^5$~yr.  This is much greater than the sound-crossing time-scale 
($\sim10^2$~yr), so we see immediately that each cloud responds 
adiabatically to pressure perturbations and consequently dynamical 
stability is assured.  Thermal stability is another matter.

Because the heating of each cloud is largely independent of
its temperature, the radiative cooling must actually
{\it decrease,\/} with increasing $T$, if it is to be
thermally stable. If it were otherwise, the following
scenario would take place. Suppose the cloud contracts slightly
from an initial equilibrium condition, then its temperature
increases, consequently it radiates more efficiently,
but the heating rate remains the same, so cooling then
outstrips heating and this causes further contraction.
Evidently this contraction can continue without limit
under the stated circumstances and such a cloud would
be thermally unstable, collapsing (or expanding) on a
thermal time-scale. As this is much less than the Hubble
time, this is not an acceptable model. The difficulty we
then face is that, in order to construct a thermally stable
model, we need to identify a radiative cooling process
which becomes {\it less\/} effective at higher temperatures.

While it is possible for this circumstance to arise, e.g. by
virtue of a single spectral line becoming optically thick, it
is a highly anomalous situation. In the particular case
of the hypothesised cold, dense clouds there happens to
be a remarkably simple solution to this conundrum (Wardle
\& Walker 1999). Conditions within the clouds are close
to those required for the precipitation of solid hydrogen
(Pfenniger \& Combes 1994), and such particles efficiently
cool the gas via their thermal continuum
radiation. If this process dominates
the radiative cooling of the clouds, then they can
be thermally stable because an increase in temperature
rapidly destroys the coolant, leading to less efficient
cooling at higher temperatures. This concept is examined
in detail by Wardle \& Walker (1999).

\subsection{Cloud survival}

Beyond just the stability of the clouds, we need to understand
their survival in the Galaxy for
billions of years. This issue was first considered by Gerhard
\& Silk (1996), who arrived at simple criteria which the clouds
must satisfy if they are not to be destroyed by evaporation or
by collisions. Gerhard \& Silk (1996) found the latter to be
the more restrictive condition, implying a lower bound on the
column density of individual clouds: $N\gta4\times10^{24}\;{\rm cm^{-2}}$.
This constraint can be tightened quite considerably by
going beyond just order-of-magnitude estimates and examining
the evolution, under the influence of collisions, of a
model halo.

Starting from a singular isothermal sphere, Walker (1999b) found
that the dark halo developed a core of constant density, with
the core radius increasing as a function of time, as a result of
destructive collisions between clouds. In this model the size of
the core, $r_c$, is a function of the halo velocity dispersion,
$\sigma$ -- large velocity dispersions imply high collision rates
-- with $r_c\propto\sigma^{3/2}$. There is some evidence that dark
halos do indeed possess finite cores with the size of the core
increasing as $\sigma^{3/2}$ (Kormendy 1990).

Collisions between clouds cause shock heating of their constituent
material; these shocks typically dissociate the molecular gas and
unbind the clouds, their material subsequently being assimilated
into the ISM of their host galaxy. Walker \& Wardle (1998b) pointed
out that the Galactic High Velocity Clouds
(HVCs; Wakker \& van~Woerden 1997) might plausibly be
identified with post-collision gas which has not yet been assimilated
into the Galactic disk. If this identification is correct, the fact
that HVCs contain metals is of particular interest because our
naive expectation is that the composition of these clouds
reflects the nucleosynthetic yields of the Big Bang. This is
entirely unconventional, of course, but is in accord with the
hypothesis that the clouds predate nucleosynthesis and introduce
inhomogeneity at that epoch (\S\S5.1,5.2; Applegate \& Hogan 1985).

In due course stars may form from the diffuse gas released by
collisions, but either way the material is part of the visible pool
of matter and consequently the visible mass within a dark halo
can be predicted. Walker (1999b) discovered that
data published by Broeils (1992) are in good agreement with
this model; matching the theory to the data requires only that the
surface density of individual clouds is $\Sigma\simeq140\;
{\rm g\,cm^{-2}}$ (assuming the age of the Universe to be 10~Gyr).
Moreover it seems very likely that this relationship between
visible galaxy mass and halo velocity dispersion,
$M_{vis}\propto\sigma^{7/2}$, underlies the well-known Tully-Fisher
relation between galaxy luminosity and velocity dispersion.
This is a remarkable success for such a simplistic model, and
this result assumes particular importance because it occurs
in the arena where we find the strongest evidence for dark matter,
i.e. the dynamics of spiral galaxies.

With the above estimate for the mean cloud surface density we
can immediately compute the sky-covering fraction for the
clouds: $f_{coll}\simeq2\times10^{-4}$ (Walker 1999b). This
is considerably smaller than the estimate based on ESEs
($f_{ESE}\sim5\times10^{-3}$; Fiedler et al 1994), and
if we wish to reconcile these values it seems necessary to
contemplate photo-ionised winds arising at radii
$(f_{ESE}/f_{coll})^{1/2}\sim5$ times larger than the
underlying hydrostatic clouds. Presumably, if both models
are valid, this difference implies that a neutral wind
transports mass out to several cloud radii before it
becomes ionised. Thus since Walker \& Wardle (1998a)
estimate an inner radius of order 2~AU for the photo-ionised
wind, we can estimate the underlying cloud radii as roughly
0.4~AU$=6\times10^{12}\;{\rm cm}$. In combination with the
estimated mean surface density, this implies that a
characteristic mass for the individual clouds is
$M\sim10^{-5}\;{\rm M_\odot}$. (A cloud with
this mass/radius ratio has a virial temperature of
a few Kelvin, cf. \S5.3.) As the line of
reasoning which leads to this figure is not yet secure,
the estimate should be treated with some caution.

A further interesting aspect of collisions is that the shock-heated
gas radiates strongly, and this radiation may be detectable. In the
case of collisions in the Galactic halo one expects (unobservable)
extreme ultraviolet flashes, with accompanying optical transients
at a median magnitude of $V<23$, lasting a few days. For halos
with larger velocity dispersions the thermal radiation moves into
the X-ray band and is directly observable. Rather dramatically,
the implied X-ray luminosity for a cluster of galaxies with
$\sigma=1000\;{\rm km\,s^{-1}}$ is (assuming polytropic clouds
with n=3/2) $L_X\simeq3\times10^{44}\;{\rm erg\,s^{-1}}$,
comparable to what is actually observed from such clusters.
Because this emission is thermal radiation from gas at roughly
the virial temperature, it is difficult to distinguish it from
thermal emission by {\it diffuse,\/} hot gas in the cluster (which
is the standard interpretation of the observed emission). These
issues are discussed by Walker (1999c).

\section{Observational tests}

Having established that cold clouds could make up the dark matter,
i.e. there is no evident inconsistency with observations, we can
turn to the task of isolating some critical tests of
the picture. Unlike the non-baryonic dark matter candidates, cold
clouds interact with their environment in a rich variety of ways,
admitting direct tests of the theory.

\subsection{Lensing phenomena}

A clear prediction of the model is that cosmologically distant
sources should be strongly gravitationally lensed by intervening
clouds. This possibility can be investigated relatively cleanly if
one studies a large sample of quasars located behind low-redshift
over-densities such as galaxy halos or clusters of galaxies
(Walker 1999a; Walker \& Ireland 1995; Tadros, Warren \& Hewett 1998).
In a collaborative effort led by Robert Smith (ANU), we are
working towards this goal, making use of the large sample of
objects identified in the 2dF Quasar Survey (Boyle et al 1996).

For clouds located in the Galactic halo, gravitational lensing
is insignificant, but {\it gas\/} lensing may introduce measurable
flux changes (Draine 1998). In the absence of a reliable structural
model for the clouds there is great uncertainty in the predicted
(de)magnification, making it difficult to use as a test. Nevertheless
the requisite data have already been accumulated in the search for
gravitational lensing events against LMC stars, and these data could
usefully be searched for gas lensing events. Such a search should
employ different criteria to those used to find gravitational microlensing
(e.g. Alcock et al 1998): for gas lensing quite large differences
in magnification can occur for different colours, and the light
curves can be quite dissimilar to those for gravitational lensing
by a compact object. We note also that the arguments presented
in \S5.4, suggesting a relatively small radius for the hydrostatic
cloud, imply (high magnification) event time-scales of less than
a day (cf. Draine 1998), and light-curves which are profoundly
influenced by source size.

Plasma lensing (ESEs and related phenomena) is an area where
further observational work would be extremely helpful. Pulsars
offer the best targets for such work: they are much more informative
than quasars in respect of the properties of the lens; multiple
imaging is more common at low frequencies, where pulsars are
brightest; and pulsars are very small in angular size, admitting
sensitivity to distant lenses. Even rather basic information about
the properties of the plasma lenses could discriminate between the
various models. For example: if the lenses could be shown to be
approximately axisymmetric and possessing an off-axis peak in
electron column density, this evidence would very strongly favour
the cold-cloud model. We also note that a 21~cm absorption line
should arise during an ESE, but the strength of this line is
difficult to predict because it depends on the tiny fraction
of atomic hydrogen present within the cloud.

\subsection{Extinction events}

Throughout the far UV and X-ray bands the clouds must be
completely opaque, implying that a small fraction of all compact
extragalactic sources should be extinguished in these wavebands.
However, these bands require space-based instrumentation and
intensive monitoring experiments are consequently difficult to
pursue. The best prospect for discovering clouds through their
extinction therefore seems to be an accurate monitoring program at the
blue end of the optical band, looking for Rayleigh scattering by
the H$_2$. Modelling the cloud-cloud collision
process (\S5.4; Walker 1999b) provides an estimate of the mean cloud
surface density $\Sigma\simeq140\;{\rm g\,cm^{-2}}$, which translates
to a mean column density in H$_2$ of $3.1\times10^{25}\;{\rm cm^{-2}}$.
Using the scattering cross-section of Dalgarno \& Williams (1965)
this implies an average extinction of $\Delta B\simeq0.073$~magnitudes,
while $\Delta V\simeq0.032$ and $\Delta R\simeq0.012$. (Gas lensing
[\S\S4.4.2, 6.1] also affects the received flux; however, for most
of the time during a gas lensing event the received flux is lower
than if the lens were absent, so that gas lensing typically reinforces
the effects of extinction.) At shorter wavelengths, scattering
quickly becomes a large effect: 0.25~mag at 336~nm, and 0.85~mag
at 255~nm. Extinction events should last only a week or two, and
should be manifest in a fraction $\simeq2\times10^{-4}$ of
Magellanic Cloud stars, say, at any one time.

\subsection{Local H$\alpha$ sources}

By virtue of their small mass, the nearest dark clouds ought to be
quite close to the sun -- perhaps within 0.1~pc -- and it may be
possible to detect these objects through their H$\alpha$ emission.
In the solar neighbourhood the mean intensity of ionising photons
appears to be very low (Vallerga \& Welsh 1995) and will not lead to
a detectable emission measure, even for a cloud which is so close
that it is resolved by the telescope. However, the cosmic rays
which pass through the cloud create some ionisation throughout
its volume, and a small fraction of these ionisations lead to
the production of H$\alpha$ photons.

More specifically, cosmic-ray ionisation of He gives He$^+$ which
reacts with H$_2$ to give He, H and H$^+$, the last of which
recombines with an electron, yielding emergent Balmer photons.
(By contrast, ionisation of H$_2$ leads to the formation of H$_2^+$,
which reacts with H$_2$ to give H$_3^+$; this subsequently recombines
with an electron to yield H$_2$ and H.) Assuming that roughly 60\% of
H$^+$ recombinations yield an H$\alpha$ photon -- as for Case B
conditions -- and adopting an ionisation rate of $3\times10^{-17}\;{\rm s^{-1}}$
(e.g. Webber 1998), the implied H$\alpha$ luminosity is $2\times10^{35}M_{-4}
\;{\rm s^{-1}}$, where $M_{-4}$ is the cloud mass in units of 
$10^{-4}\mathrm{M}_\odot$. Now from the results of Walker (1999b) we can
infer a local cloud density of $80/M_{-4}\;{\rm pc^{-3}}$, so if we
survey a solid angle of $\omega$~sr, the brightest cloud within
the survey area is expected to have a flux of $1.5\times10^{-2}
M_{-4}^{1/3}\omega^{2/3}\;{\rm cm^{-2}\,s^{-1}}$.

By good fortune it happens that the Anglo-Australian Observatory
has recently commissioned an H$\alpha$ filter for use with its
Schmidt telescope. In three hours this combination reaches a depth
approximately equivalent to $R=21$ (Parker 1998), and using
the magnitude-flux transformation of Bessell (1983) we find that
this corresponds to $2.0\times10^{-4}\;{\rm cm^{-2}\,s^{-1}}$.
In collaboration with Quentin Parker (ROE) and Mike Irwin (IoA),
we are currently using this telescope/filter combination to
search for local examples of the cold cloud population. The
technique we are using is to observe eight high-latitude fields
(total $\omega=3.1\times10^{-2}\;{\rm sr}$), in both H$\alpha$
and $R$ to similar depth, then re-observing these fields one
year later in order to identify any high proper-motion
emission-line sources. Sciama (1999) has pointed out that the
local cloud population should also be detectable with future
satellite missions (MAP and Planck) designed to study the Cosmic
Microwave Background. These missions are, however, some years away.

\subsection{Collisions}

The theory described in \S5.4 implies that optical transients
will arise from cloud-cloud collisions occurring within the halo
of our Galaxy (Walker 1999c). We estimate that the median of the
distribution of peak visual magnitudes will be $V<23$; the total
event rate will be roughly $0.7/M_{-4}\;{\rm deg^{-2}\,yr^{-1}}$;
and typical durations will be a few days. The estimated magnitude
assumes that the internal density profile is approximately that of
a convective ($n=3/2$) polytrope; more centrally concentrated
profiles result in fainter (median) transients (e.g. fainter by
3.6~mag if $n=3$). Although the properties of the transients are at
present only crudely predicted by the theory, we know of no similar
physical phenomenon and it seems unlikely that, if discovered,
the origin of such events would be incorrectly attributed.
Discovery would require a fairly deep, wide-area monitoring
program with daily visits and rapid spectroscopic follow-up.
Because the radiating gas reaches temperatures of approximately
$5\times10^5$~K, optical spectra will presumably display 
emission lines superimposed on a continuum dominated by
free-free emission and ${\rm He^{++}}$ recombination.

\subsection{$\gamma$-ray background}

A direct view of the bulk of the clouds is afforded by their
$\gamma$-ray emission, and it would be helpful to study the
spectral properties and angular distribution of the $\gamma$-ray
background. Unfortunately there appears to be no immediate
prospect for such studies. The Compton GRO satellite was able
to detect the Galactic component of the high energy $\gamma$-ray
background (Dixon et al 1998), but lacked the sensitivity
necessary to conduct studies on angular scales much less than
a radian. Future gamma-ray missions -- notably NASA's GLAST
satellite (http://glast.gsfc.nasa.gov) -- will have much greater
sensitivity than CGRO, but are currently some years away from
realisation. In the meantime it would be profitable to
calculate the spectrum which is expected from cosmic-ray
interactions with clouds of high column density. This may
be relevant to the existing EGRET data on diffuse emission,
whose spectra are not understood (Mori 1997).

\section{Theoretical outlook}

In addition to observations designed to test the proposed
picture, there is a need to develop the model more fully.
The most blatant deficiency of the current picture is that
the origin of the putative clouds is unspecified. As
described in \S5.2, there are some hints that the clouds
formed in the very early Universe, prior to synthesis of
the light elements, and this scenario could usefully be
pursued theoretically. Unfortunately the physics appropriate
to this era involves non-perturbative QCD and is currently
unclear, so definitive investigations are not possible
at present.

The physics relevant to the present day clouds is not
subject to these difficulties, and theoretical models
of their structure can be constructed. For the central,
hydrostatic core it is especially important to determine
the density profile; this profile affects the
typical kinetic energy dissipated during collisions, as
well as the gas- and gravitational-lensing properties
of the clouds. Similarly, a better understanding of the
structure of the photo-evaporated wind would improve the
predictive power of radio-wave lensing calculations --- a
critical area at present, given that radio-wave lensing
phenomena currently provide much of our information on the
clouds' properties.

It would be very useful to acquire a better picture of
how the clouds evolved, from some (assumed) initial
conditions right up to the present day. While the correct
initial conditions are unknown at present, because the
origin of the clouds is murky, such evolutionary
calculations would help to clarify which formation scenarios
are plausible. The properties of the cloud population
prior to the epoch of recombination are particularly
interesting, because these can be related to the measurements
of Cosmic Microwave Background anisotropies. If the clouds
are sufficiently dense that they decouple from the CMB well
before recombination, then the development of large-scale
structure in the Universe will presumably be qualitatively
similar to that of the Cold Dark Matter (CDM) model, as the
clouds are expected to be dynamically cold at decoupling.
Some of the theoretical ``machinery'' currently used with CDM
-- notably the numerical simulations of structure formation
-- could then be usefully applied to the cold cloud picture.
As described in \S5.4, however, the effects of cloud-cloud
collisions on the dark matter distribution function would
have to be incorporated in any simulations.

\section{Summary}

Extreme scattering events reveal the presence
of compact, high pressure regions of ionised gas in the Galaxy;
these are best explained in terms of photo-ionisation of cold
clouds, but these clouds must contribute substantially to the
Galactic dark matter. This interpretation involves no incompatibility
with other data and most of its challenges are theoretical
--- i.e. the need to understand the origin, structure and
evolution of the clouds. Model predictions to date have
been in remarkable accord with observation;
in particular the Tully-Fisher relation for spiral galaxies
emerges as a natural consequence of the model.

The most controversial arena for the cold cloud model at present
is cosmology, where the view that most of the dark matter is
non-baryonic is usually rather firmly held. Nevertheless
one can sensibly contemplate a purely baryonic universe
if the baryons are inhomogeneously distributed at nucleosynthesis,
and these inhomogeneities may be identified with the proto-clouds
themselves. Consistency with the observed large-scale CMB
anisotropies then requires either isothermal primordial density
fluctuations or, if the fluctuations are adiabatic, early
decoupling of the clouds from the CMB.

The cold cloud model still requires a great deal of theoretical
development and observational investigation; in particular we
need a definitive observational test of the existence of the
putative clouds.





\section*{Acknowledgements}

The authors are indebted to many colleagues for the insights they
have shared with us on the topics described herein. In particular
we wish to thank: James Binney; Roger Blandford; Leo Blitz; Bruce Draine;
Ron Ekers; Eric Feigelson; Ken Freeman; Gordon Garmire; Mike Hawkins;
Peter Kalberla; Ulrich Mebold; Bohdan Paczy\'nski; Daniel Pfenniger; Sterl Phinney;
Barney Rickett; Dennis Sciama; Lister Staveley-Smith; and Ravi Subrahmanyan.


\section*{References}

\reference Alcock~C. et~al. 1997 ApJ 486, 697

\reference Alcock~C. et~al. 1998 ApJL 499, L9

\reference Applegate~J.H. \& Hogan~C.J. 1985 Phys. Rev. D 31, 3037

\reference Ashman~K.M. 1992 PASP 104, 1109

\reference Bessell~M.S. 1983 PASP 95, 480

\reference Binney~J. \& Tremaine~S. 1987 ``Galactic Dynamics'' (Princeton
univ. Press: Princeton)

\reference Bland-Hawthorn~J. \& Maloney~P.R. 1997 PASA 14, 59

\reference Bloemen~H. 1989 ARAA 27, 469

\reference Boyle~B.J., Smith~R.J., Shanks~T., Croom~S.M. \& Miller~L.
1998 proc. IAU Symp. 183, Cosmological parameters and the evolution of the
Universe

\reference Broeils~A. 1992 ``Dark and visible matter in spiral galaxies''
PhD Thesis (Groningen)

\reference Carr~B. 1994 ARAA 32, 531

\reference Clegg~A.W., Chernoff~D.F. \& Cordes~J.M. 1988 in Radio Wave
Scattering in the
 Interstellar Medium, ed. J.M.~Cordes, B.J.~Rickett \& D.C.~Backer (AIP:
New York) 174

\reference Combes~F. \& Pfenniger~D. 1997 A\&A 327, 453

\reference Cravens~T.E. \& Dalgarno~A. 1978 ApJ 219, 750

\reference Dalgarno~A. \& Williams~D.A. 1965 Proc. Phys. Soc. Lond. 85, 685

\reference Davis~M., Efstathiou~G.P., Frenk~C.S. \& White~S.D.M. 1985 ApJ
292, 371

\reference de Paolis~F., Ingrosso~G., Jetzer~Ph. \& Roncadelli~M. 1995
Phys. Rev. Lett. 74, 14

\reference Dixon~D.D. et al 1998 New Ast. 3, 539

\reference Dove~J.B. \& Shull~J.M. 1994 ApJ 430, 222

\reference Draine~B.T. 1998 ApJL 509, L41

\reference Dyson~J.E 1968 Ap\&SS 1, 388

\reference Fiedler~R.L., Dennison~B., Johnston~K.J. \& Hewish~A. 1987 Nat
326, 675

\reference Fiedler~R.L., Johnston~K.J., Waltman~E.B. \& Simon~R.S. 1994
ApJ 430, 581

\reference Fixsen~D.J. et al 1994 ApJ 420, 457

\reference Gerhard~O. \& Silk~J. 1996 ApJ 472, 34

\reference Hawkins~M.R.S. 1993 Nat 366, 242

\reference Hawkins~M.R.S. 1996 MNRAS 278, 787

\reference Henriksen~R.N. \& Widrow~L.M. 1995 ApJ 441, 70

\reference Hogan~C. 1978 MNRAS 185, 889

\reference Hogan~C.J. 1993 ApJL 415, L63

\reference Hoyle~F. 1953 ApJ 118, 513

\reference Kormendy~J. 1990 ``Evolution of the universe of galaxies'' ed.
R.~G.~Kron ASP Conf. Ser. 10, 33

\reference Kurki-Suonio~H., Matzner~R.A., Olive~K.A. \& Schramm~D.N. 1990 ApJ 353, 406

\reference Lagache~G., Abergel~A., Boulanger~F. \& Puget~J.-L. 1998 A\&A 333, 709

\reference Mori~M. 1997 ApJ 478, 225

\reference Paczy\'nski~B. 1986 ApJ 304, 1

\reference Palla~F., Salpeter~E.E. \& Stahler~S.W. 1983 ApJ 271, 632

\reference Parker~Q. 1998 Personal Communication

\reference Peebles~P.J.E. 1993 ``Principles of physical cosmology''
(Princeton Univ. Press: Princeton)

\reference Pfenniger~D., Combes~F. \& Martinet~L. 1994 A\&A 285, 79 (PCM94)

\reference Pfenniger~D. \& Combes~F. 1994 A\&A 285, 94

\reference Press~W.H. \& Gunn~J.E. 1973 ApJ 185, 397

\reference Reach~W.T. et al 1995 ApJ 451, 188

\reference Rees~M.J. 1976 MNRAS 176, 483

\reference Reynolds~R.J. 1992 ApJ 392, L35

\reference Rickett~B.J. 1990 ARAA 28, 561

\reference Romani~R., Blandford~R.~D. \& Cordes~J.~M. 1987 Nat 328, 324

\reference Sahu~K. 1994 Nat 370, 275

\reference Schild~R.E. 1996 ApJ 464, 125

\reference Sciama~D.W. 1999 MNRAS (Submitted: astro-ph/9906159)

\reference Schramm~D.N. \& Turner~M.S. 1998 Rev. Mod. Phys. 70, 303

\reference Smoot~G.F. et al 1992 ApJL 396, L1

\reference Spooner~N.J.C. 1997 Proceedings of the First International
Workshop on Identification of Dark~Matter (World~Scientific: Singapore)

\reference Tadros~H., Warren~S. \& Hewett~P. 1998 New Ast. Rev. 42, 115

\reference Trimble~V. 1987 ARAA 25, 425

\reference Vallerga~J.V. \& Welsh~B.Y. 1995 ApJ 444, 702

\reference van~den~Bergh~S. 1999 PASP (In press; astro-ph/9904251)

\reference Wakker~B. \& van~Woerden~H.~1997 ARAA 35, 217

\reference Walker~M.A. 1999a MNRAS (In Press; astro-ph/9806196)

\reference Walker~M.A. 1999b MNRAS (In Press; astro-ph/9807236)

\reference Walker~M.A. 1999c (In preparation)

\reference Walker~M.A. \& Ireland~P.M. 1995 MNRAS 275, L41

\reference Walker~M. \& Wardle~M. 1998a ApJL 498, L125

\reference Walker~M. \& Wardle~M. 1999 ``Proc. Stromlo Workshop on High
Velocity Clouds'' (eds. B.K.~Gibson \& M.E.~Putman) ASP Conf. Ser. 166, p269

\reference Wardle~M. \& Walker~M. 1999 ApJL (Submitted: astro-ph/9907023)

\reference Webber~W.R. 1998 ApJ 506, 329

\reference Webber~W.R., Lee~M.A. \& Gupta~M. 1992 ApJ 390, 96

\reference Zwicky~F. 1933 Helv. Phys. Acta 6, 110







\end{document}